\documentclass[10pt,pra,twocolumn,showpacs,floatfix]{revtex4}

\usepackage{graphicx}
\usepackage{amssymb}
\usepackage{times}
\usepackage{amsmath}
\usepackage{amsthm}
\usepackage{subfigure}
\usepackage{float}
\usepackage[breaklinks]{hyperref}
\input epsf.tex
\usepackage{graphicx}
\usepackage{amsthm}

\begin{document}

\title{Toward Ultra-high Sensitivity in Weak Value Amplification}% Force line breaks with\\

\author{Jingzheng Huang\footnote{Email: jzhuang1983@sjtu.edu.cn}, Yanjia Li, Chen Fang, Hongjing Li, and Guihua Zeng\footnote{Email: ghzeng@sjtu.edu.cn}}
\affiliation
{State Key Laboratory of Advanced Optical Communication Systems and Networks, 
	$\&$ Center of Quantum Sensing and Information Processing, 
	Shanghai JiaoTong University, Shanghai 200240, China}
%\date{\today}

\begin{abstract}
Achieving higher sensitivity is an earnest purpose for precision metrology. As a response to this goal, the weak value amplification approach has been developed for measuring ultra-small physical effects, realizing sensitivity that had never been reached before. Encouraged by the successes, many efforts have been devoted to obtain ultimate sensitivity of weak value amplification. However, the benefit would be easily compromised in practice, because the cost of significant reduction on signal intensity leads to an ultra-low signal-to-noise ratio. In this work, we bridge this gap by proposing an alternative weak value amplification approach, which provides sensitivity several orders of magnitude higher than the standard approach while being compatible with practical imperfections. In the proof-of-principle experiment of measuring longitudinal phase change in time-domain, sensitivity up to $5\times10^{-4}$ attosecond is exemplified. Our approach can be applied to measure other small parameters with extremely high sensitivity, providing a new method for future precision metrology.
\end{abstract}

\maketitle

\section{Introduction} 
Since the concept of weak value amplification (WVA) was proposed in 1988\cite{Aharonov1988,Aharonov1990}, its power has been widely demonstrated in numerous applications\cite{Hulet1991,Aharonov2002,PL2010,Ludeen2011,Jordan2014,Kim2018}. 
In particular for the demand of precision metrology, WVA provides extremely high sensitivity for observing many ultra-small physical effects that have never been realized by other techniques before\cite{Dressel2014}.
For instance, the first observation of the spin Hall effect of light was performed by using WVA technique, in which a sensitivity to displacement of ~1 $\AA$ was achieved\cite{Hosten2008}. Recently, this record has been improved to the level of femtometer\cite{Dixon2009,Martinez-Rincon2017}. More examples could be found in the applications on measuring other parameters with high sensitivities, e.g. velocity@400fm/s\cite{Viza2013}, non-linearity@single-photon level\cite{Chen2018}, temperature@$3\times10^{-6o}C$\cite{Egan2012,Li2018}, optical rotation@$1.88\times10^{-5o}$\cite{Xu2018a}, time-domain longitudinal phase change@sub-attosecond level\cite{Xu2013,Fang2016}, etc. These applications employ the standard WVA approach, in which only a tiny portion of the signal surviving  post-selection are collected and measured\cite{Kofman2012}. Encouraged by these huge successes, many efforts were made to pursue the ultimate sensitivity. Impressively, it was pointed out by Ref.\cite{Zhang2016} that there exists an ultimate sensitive working area, with sensitivity several orders of magnitude higher than the standard approach\cite{Brunner2010,Li2011}. The cost of it is reducing the post-selection probability by the same orders of magnitude. In practice, smaller signal intensity indicates lower signal-to-noise ratio when facing same amount of background noises, thus the benefit gained by sensitivity enhancement would be completely compromised\cite{Xu2018b}, and its implementation is practically challenging. Recently, another approach using joint detection was proposed\cite{Strubi2013}. In contrast to the previous proposals, this approach exploits the full information contained in all outputs, therefore larger signal intensity and higher signal-to-noise ratio can be achieved\cite{Martinez-Rincon2016,Fang2016}. However, the sensitivity of this approach is by half of the standard approach\cite{Martinez-Rincon2016}. 

Here, we propose a novel approach called dual weak value amplification (DWVA), with sensitivity several orders of magnitude higher than the standard approach without losing signal intensity. Our approach merges the advantages of using optimal working area\cite{Zhang2016} and joint detection\cite{Strubi2013}, thus simultaneously harvests the ultra-high sensitivity and high signal-to-noise ratio. Moreover, distinct from the previous proposals\cite{Howell2010,Brunner2010,Li2011}, our approach is successfully demonstrated with a non-Gaussian initial pointer state, which implies that it could be more compatible with the practical imperfections. For demonstration, we apply our approach to perform a proof-of-principle experiment, where a tiny optical longitudinal phase change is converted to a large mean wavelength shift. Taking the advantages of high enough signal intensity and tolerable to light source imperfections, we achieve a mean wavelength shift rate of over 20nm per attosecond, at a level that has never been achieved before. Besides longitudinal phase change measurement, our approach can be extended to applications on other metrological tasks that require extremely high sensitivity.

\section{Theory} 
Let us begin with considering a weak value amplification scenario involving a two-level system and a pointer with continuous degree of freedom, where the initial state of system is $|\psi_{in}\rangle$ and the initial spectrum of pointer with variable $p$ is $f(p)$. And then, system and point are weakly interacted, the process is described by a unitary operator\cite{Nielsen2001} 
\begin{equation}\label{eq:U_int}
\hat{U}_{int}(g) = \exp[i g \hat{A}\otimes\hat{P}] = \cos(g\hat{p})\hat{\mathbb{I}} - i\sin(g\hat{p})\hat{A},
\end{equation}
where $\hat{A}$ acts on the system with eigenvalues of $\pm 1$, $\hat{p}$ acts on the pointer, and $g$ indicates the coupling strength. After the interaction, the evolved state is projected to two final states $|\psi_{f1}\rangle$ and $|\psi_{f2}\rangle$. In contrast to the previous WVA proposals, the initial and final system states of DWVA are respectively modulated by:
\begin{equation}\label{eq:in_final}
\begin{array}{lll}
|\psi_{in}\rangle &= (e^{-i(1-\frac{p}{p_0})\epsilon}|+1\rangle_s + e^{i(1-\frac{p}{p_0})\epsilon}|-1\rangle_s)/\sqrt{2},\\
|\psi_{f1}\rangle &= (|+1\rangle_s + i|-1\rangle_s)/\sqrt{2},\\
|\psi_{f2}\rangle &= (|+1\rangle_s - i|-1\rangle_s)/\sqrt{2},
\end{array}
\end{equation}
where $\epsilon \ll 1$ is a small $p$-independent phase shift, $p_0$ is the mean value of $p$ calculated by $p_0=\int pf(p)dp$. Here, $p_0\gg1$ and $p_0g\ll1$ are required. In addition, we assume that the variance of $f(p)$ (denoted as $\sigma_p$) is much smaller than $p_0$. Accordingly, the weak values\cite{Aharonov1990} corresponding to each final state, which are defined by $A_{wk}=\frac{\langle\psi_{fk}|\hat{A}|\psi_{in}\rangle}{\langle\psi_{fk}|\psi_{in}\rangle}$ with $k=1,2$, can be derived by:
\begin{equation}
\begin{array}{lll}\nonumber
A_{w1}(p) = i\frac{1+\sin[2(1-\frac{p}{p_0})\epsilon]}{\cos[2(1-\frac{p}{p_0})\epsilon]}, 
A_{w2}(p) = -i\frac{1-\sin[2(1-\frac{p}{p_0})\epsilon]}{\cos[2(1-\frac{p}{p_0})\epsilon]}.
\end{array}
\end{equation}
Meanwhile, the pointer spectrum corresponding to different final system states are given by $|\langle\psi_{fk}|\psi_{in}\rangle|^2\zeta_k(g,p)P_0(p)$ with $k=1,2$, and we get\cite{Fang2016}:
\begin{equation}\label{eq:P_i}
\begin{array}{lll}
P_1(p) &\approx \frac{1}{2}[1-\sin(2(1-\frac{p}{p_0})\epsilon)]\{1+2gpIm[A_{w1}(p)]\}P_0(p),\\
P_2(p) &\approx \frac{1}{2}[1+\sin(2(1-\frac{p}{p_0})\epsilon)]\{1+2gpIm[A_{w2}(p)]\}P_0(p),
\end{array}
\end{equation}
where $P_0=|f_0(p)|^2$ is the initial pointer spectrum, 
$$\zeta_k(g,p)\equiv\cos^2(gp)+\sin^2(gp)|A_{wk}(p)|^2+\sin(2pg)ImA_{wk}(p)$$
is the component that shapes the final pointer spectrum. The approximation of Eq.(\ref{eq:P_i}) holds to first-order. Afterward, the data process starts from calculating the subtraction of $P_1$ and $P_2$: $\Delta P(p) = P_1(p) - P_2(p) \approx 2[-(1-\frac{p}{p_0})\epsilon+gp]P_0(p)$ and its square:
\begin{equation}\label{eq:deltaP}
[\Delta P(p)]^2 = 4[(1-\frac{p}{p_0})\epsilon-gp]^2P^2_0(p).
\end{equation}

To derive our main results, we treat the normalized form of Eq.(\ref{eq:deltaP}) as a new probability distribution, and apply it to calculate the mean value of $p$. Notice that by assuming $P_0(p)$ a Gaussian distribution, $P^2_0(p)$ is Gaussian with mean value of $p_0$ and variance of $\sigma_p/2$. Defining the mean value shift as $\delta p_{D} = \frac{\int dp p[\Delta P(p)]^2}{\int dp[\Delta P(p)]^2} - p_0$, when $g \rightarrow 0$ straightforward calculations lead to:
%
%\begin{equation}\label{eq:dp_rw}
%\begin{array}{lll}
$$\delta p_{D} \rightarrow g\times\frac{d\delta p}{dg}|_{g=0} \simeq \frac{2p_0^2}{\epsilon}g.$$
%\end{array}
%\end{equation}
%
Depending on this relation, $\delta p_D$ can be applied to estimate the parameter of interest. Meanwhile, the normalized signal intensity is calculated by:
%
%\begin{equation}\label{eq:p_rw}
$$\xi_{D} = \int dp |\Delta P(p)| \approx \frac{2}{\sqrt{\pi}}\cdot\frac{\sigma_p}{p_0}\epsilon.$$
%\end{equation}
%
It shows that the signal intensity of DWVA is significantly larger than that of the previous proposals with similar sensitivity\cite{Zhang2016}. More details would be shown as follows.

\section{Comparison to existed proposals} 
In follows, we compare DWVA with the previous approaches and illustrate its advantages. Firstly, we unified the current WVA approaches within the same theory framework. For the initial state, it can be chosen as produced state or entangled state:
\begin{equation}\nonumber%\label{eq:pre}
\begin{array}{lll}
|\Psi_{in}\rangle_{PI} = &(e^{-i\epsilon}|+1\rangle_s + e^{i\epsilon}|-1\rangle_s) \otimes \int dpf_0(p)|p\rangle_m,\\
|\Psi_{in}\rangle_{EI} = &\int dp (e^{-i(1-\frac{p}{p_0})\epsilon}|+1\rangle_s + e^{i(1-\frac{p}{p_0})\epsilon}|-1\rangle_s)\\ 
&\otimes f_0(p)|p\rangle_m,
\end{array}
\end{equation}
where the subscript $PI$ indicates "product initial-state (PI)" and $EI$ indicates "entangled initial-state (EI)", respectively. And then, the final state are chosen as the eigenstates of $\hat{X}$ (single) or $\hat{Y}$ (dual) as:
\begin{equation}\nonumber%\label{eq:final}
\begin{array}{lll}
|\psi_{f}\rangle_{SD} = (|+1\rangle_s - |-1\rangle_s)/\sqrt{2},\\
|\psi_{f}\rangle_{DD} = (|+1\rangle_s \pm i|-1\rangle_s)/\sqrt{2},
\end{array}
\end{equation}
where the subscripts $SD$ and $DD$ indicate "single detection (SD)" and "dual detection (DD)", respectively. 

There are four combinations between these initial and final states, and each combination corresponds to a weak value amplification approach, which are summarized as follows:

(1) Standard weak value amplification (SWVA)\cite{Joza2007,Brunner2010}: %PI + SD
employing PI and SD, i.e., the initial and final states are chosen as $|\Psi_{in}\rangle_{PI}$ and $|\psi_{f}\rangle_{SD}$ respectively, and the signal is given by the ones survive postselection. The mean value shift and normalized signal intensity are given by\cite{Joza2007}:\\
\begin{equation}\label{eq:swva}
\delta p_{S} \simeq \frac{2\sigma_p^2}{\epsilon}g,~~\xi_{S} \approx \epsilon^2
\end{equation}

(2) Biased weak-value-amplification (BWVA)\cite{Zhang2016}: employing EI and SD, i.e., the initial and final states are chosen as $|\Psi_{in}\rangle_{EI}$ and $|\psi_{f}\rangle_{SD}$ respectively, and the signal is given by the ones survive postselection. The mean value shift and normalized signal intensity are given by\cite{Zhang2016}:\\
\begin{equation}\label{eq:bwva}
\delta p_{B} \simeq \frac{2p_0^2}{\epsilon}g,~~\xi_{B} \approx \frac{\sigma_p^2}{2p_0^2}\epsilon^2.
\end{equation}

(3) Joint weak-value-amplification (JWVA)\cite{Strubi2013,Martinez-Rincon2016,Fang2016}: employing PI and DD, i.e., the initial and final states are chosen as $|\Psi_{in}\rangle_{PI}$ and $|\psi_{f}\rangle_{DD}$ respectively, and the signal is extracted from all outputs. The mean value shift and normalized signal intensity are given by\cite{Martinez-Rincon2016}:\\
\begin{equation}\label{eq:jwva}
\delta p_{J} \simeq \frac{\sigma_p^2}{\epsilon}g,~~\xi_{J} \approx 2\epsilon.
\end{equation}

(4) Dual weak-value-amplification (DWVA): employing EI and DD, i.e., the initial and final states are chosen as $|\Psi_{in}\rangle_{EI}$ and $|\psi_{f}\rangle_{DD}$ respectively, and the signal is extracted from all outputs. The mean value shift and normalized signal intensity are given by (derivations in this work):\\
\begin{equation}\label{eq:dwva}
\delta p_{D} \simeq \frac{2p_0^2}{\epsilon}g,~~\xi_{D} \approx \frac{2}{\sqrt{\pi}}\cdot\frac{\sigma_p}{p_0}\epsilon.
\end{equation}

\begin{figure}[!h]
	\subfigure[]{
		\begin{minipage}[t]{0.9\linewidth}
			\centering
			\includegraphics[width=8cm]{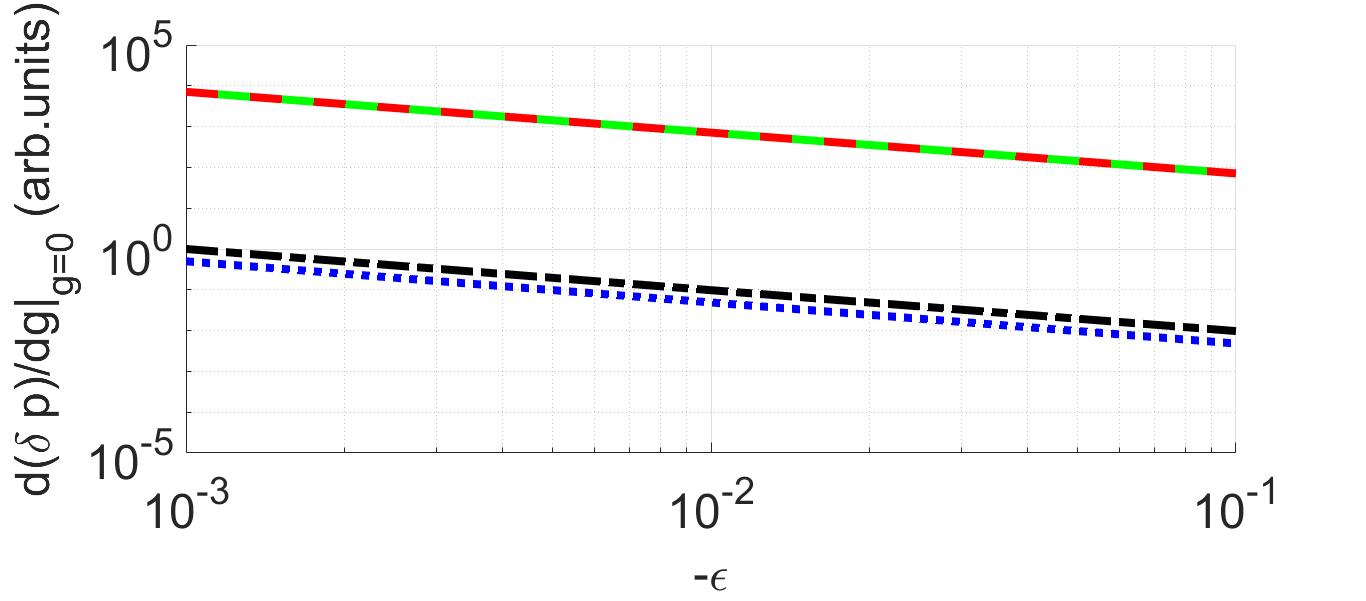}
		\end{minipage}
	}
	\subfigure[]{
		\begin{minipage}[t]{0.9\linewidth}
			\centering
			\includegraphics[width=8cm]{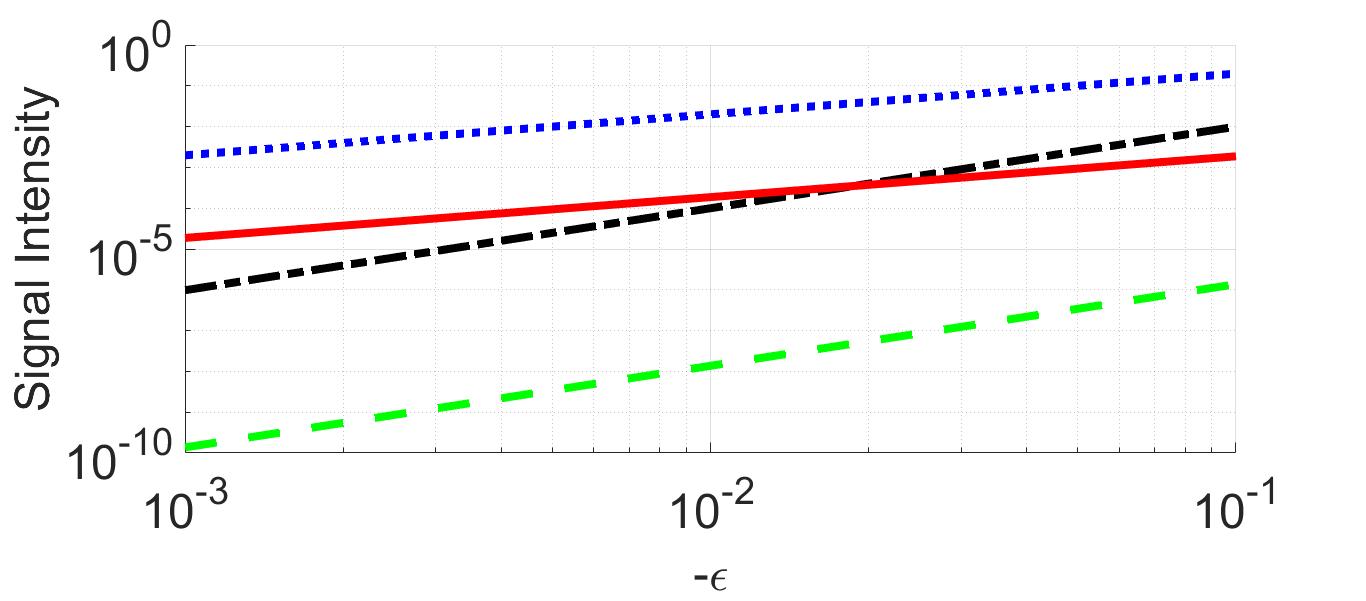}
		\end{minipage}
	}
	\caption{(Color on line) Comparison on mean value shift rates (MVSR) and signal intensities (SI) of the four WVA approaches by numerical simulations, with setting $p_0=60\sigma_p$ and the sensitivity of SWVA to $\epsilon=-0.001$ as 1. (a) Comparison on MVSR, from top to bottom: BWVA and DWVA (green-dash and red-solid, coincided), SWVA (black-dash-dot) and JWVA (blue-dot). (b) Comparison on SI, from top to bottom: JWVA (blue-dot), DWVA (red-solid), SWVA (black-dash-dot) and BWVA (green-dash).
	}\label{fig:simulation_rate_and_intensity}
\end{figure}

Intuitively, sensitivity of the WVA approaches can be defined as $\frac{d\delta p}{dg}$. By setting $p_0=60\sigma_p$, numerical simulations are performed for comparison on sensitivities and signal intensities of the four approaches. As results, Fig.\ref{fig:simulation_rate_and_intensity}(a) shows that sensitivities of DWVA and BWVA are higher than that of SWVA and JWVA by over 3 orders magnitude. On the other hand, Fig.\ref{fig:simulation_rate_and_intensity}(b) shows that the signal intensity of DWVA has the is similar to SWVA and about 4-5 orders of magnitude higher than BWVA. 
It would be worth to note that these improvements depend on the value of $\sigma_p/p_0$. A larger value of $\sigma_p/p_0$ yields higher signal intensity but lower sensitivity enhancement.

In practice, it requires the detected signal intensity high enough to ensure a certain signal-to-noise ratio. Suppose a proportion of $\xi$ with respect to the input is required. Applying Eq.(\ref{eq:swva})-(\ref{eq:dwva}), we can derive the highest achievable sensitivity (with respect to SWVA) of the four approaches. Results are summarized in Table.\ref{tab:compare}. By setting $\xi=10^{-4}$ for example, DWVA has the highest achievable sensitivity, with over 3 orders of magnitude higher than SWVA, and 1 to 2 orders of magnitude higher than BWVA and JWVA. 
\begin{table}[h]
	\centering
	\caption{Comparison on the highest achievable relative sensitivity (with respect to SWVA) of the four WVA approaches. Numerical simulation conditions: $\xi=10^{-4},p_0=60\sigma_p$.}\label{tab:compare}
	\begin{tabular}{|c|c|c|c|}
		\hline
		Approach & Sensitivity($d\delta p/dg$)       & Relative Sensitivity   & Simulation  \\\hline
		SWVA     & $2\sigma_p^2/\sqrt{\xi}$          & $1$                    & $1$        \\\hline
		BWVA     & $\sqrt{2}p_0\sigma_p/\sqrt{\xi}$  & $p_0/\sqrt{2}\sigma_p$ & $42.4$        \\\hline
		JWVA     & $2\sigma_p^2/\xi$                 & $1/\sqrt{\xi}$         & $100$        \\\hline
		DWVA     & $4p_0\sigma_p/\sqrt{\pi}\xi$      & $2p_0/\sqrt{\pi\xi}\sigma_p$ & $6770.3$  \\\hline
	\end{tabular}
\end{table}

\section{Application in longitudinal phase change measurement} 
To demonstrate the DWVA approach, we study its application in ultra-small longitudinal phase change measurement. It has been shown that WVA technique is competent to this task\cite{Brunner2010,Xu2013,Strubi2013,Fang2016,Zhang2016}, with sensitivity beyond attosecond (as) has been experimentally demonstrated\cite{Fang2016}. In theory, BWVA can achieve sensitivity of $10^{-5} as$, however this prediction has not yet been demonstrated, partly due to its extremely low output signal intensity. According to Eq.(\ref{eq:bwva}), if we typically set $\epsilon=0.08$ and use a light source with mean wavelength of $1550nm$ and line width of $25nm$, the detectable signal intensity of BWVA is with proportion of around $10^{-6}$ to the input. Suppose the input light power is 10mW, the output signal power remains only 10nW, which would be easily submerged in the background noises. 

To deal with this problem by DWVA approach, one can adopt optical polarization and angular frequency as "system" and "pointer", with replacing $p$ and $g$ by optical angular frequency $\omega$ and time-domain longitude phase change $\tau$ respectively. According to Eq.(\ref{eq:in_final}), the initial and final polarization states should be set by
\begin{equation}\label{eq:psi_in_final}
\begin{array}{lll}
|\psi_{in}\rangle = (e^{-iC}|H\rangle + e^{iC}|V\rangle)/\sqrt{2},\\
|\psi_{f1}\rangle = (|H\rangle + i|V\rangle)/\sqrt{2},\\
|\psi_{f2}\rangle = (|H\rangle - i|V\rangle)/\sqrt{2},
\end{array}
\end{equation}
where $C\equiv(1-\frac{\omega}{\omega_0})\epsilon$ with $\omega_0$ the mean value of $\omega$, $H$ and $V$ denote horizontal and vertical polarizing respectively.

\begin{figure}[!h]
	\subfigure[]{
		\begin{minipage}[t]{0.47\linewidth}
			\centering
			\includegraphics[width=4.2cm]{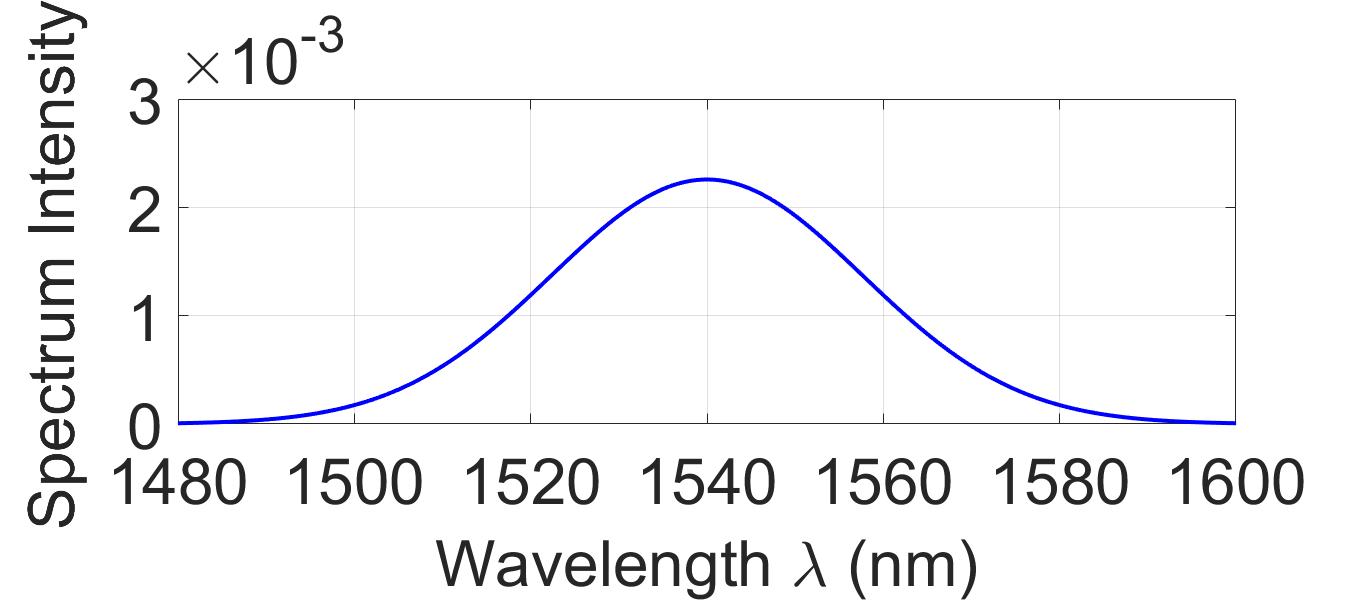}
		\end{minipage}
	}
    \subfigure[]{
    	\begin{minipage}[t]{0.47\linewidth}
    		\centering
    		\includegraphics[width=4.2cm]{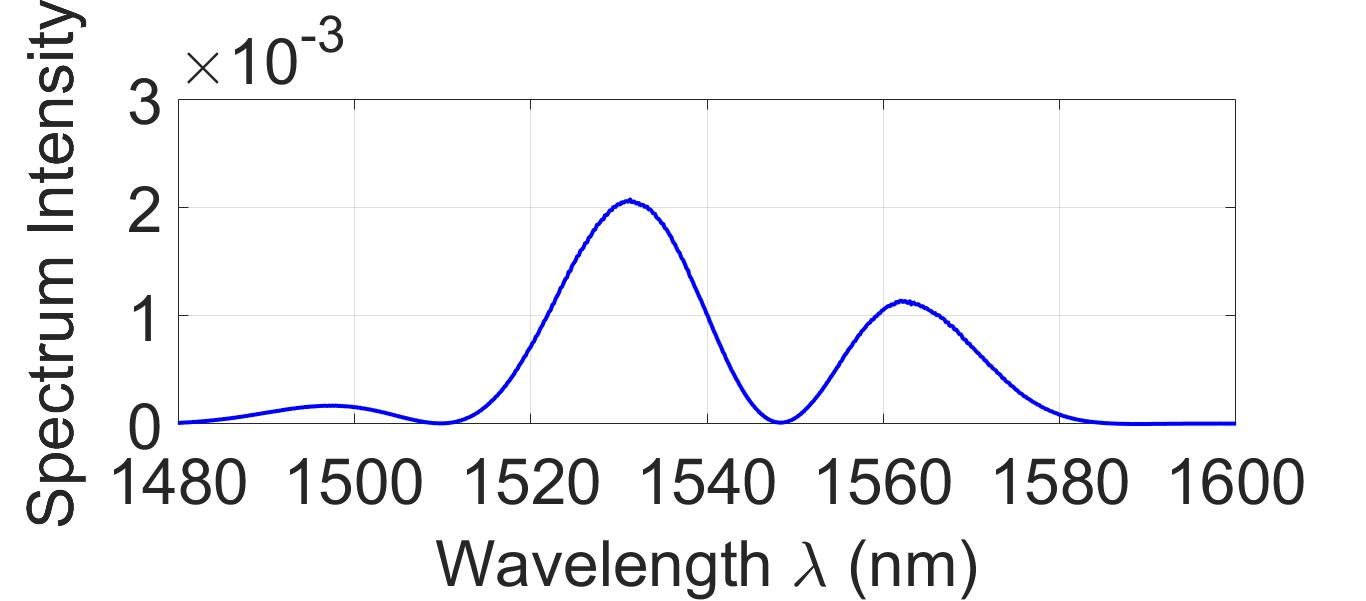}
    	\end{minipage}
    }
	\subfigure[]{
		\begin{minipage}[t]{0.47\linewidth}
			\centering
			\includegraphics[width=4.2cm]{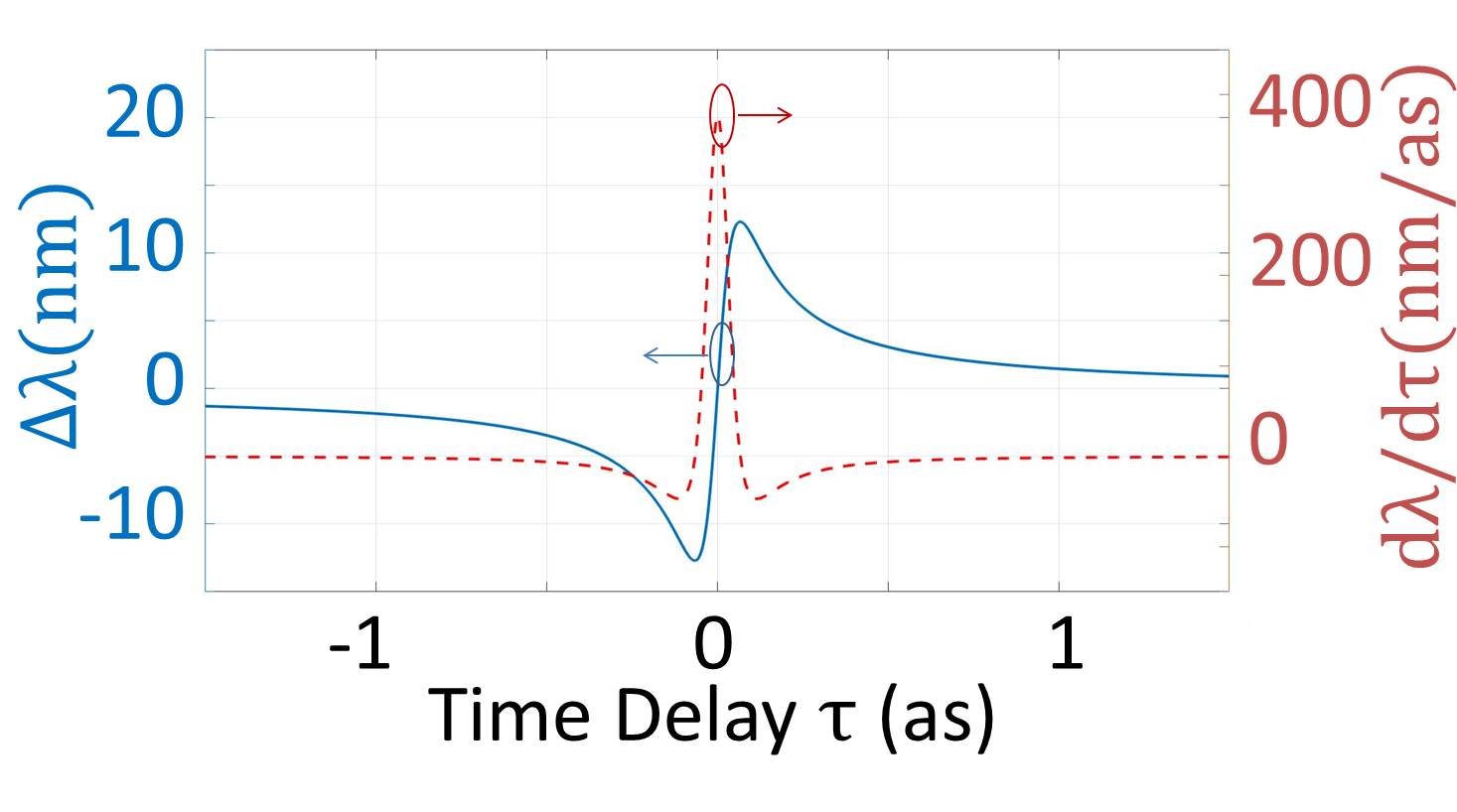}
		\end{minipage}
	}
    \subfigure[]{
    	\begin{minipage}[t]{0.47\linewidth}
    		\centering
    		\includegraphics[width=4.2cm]{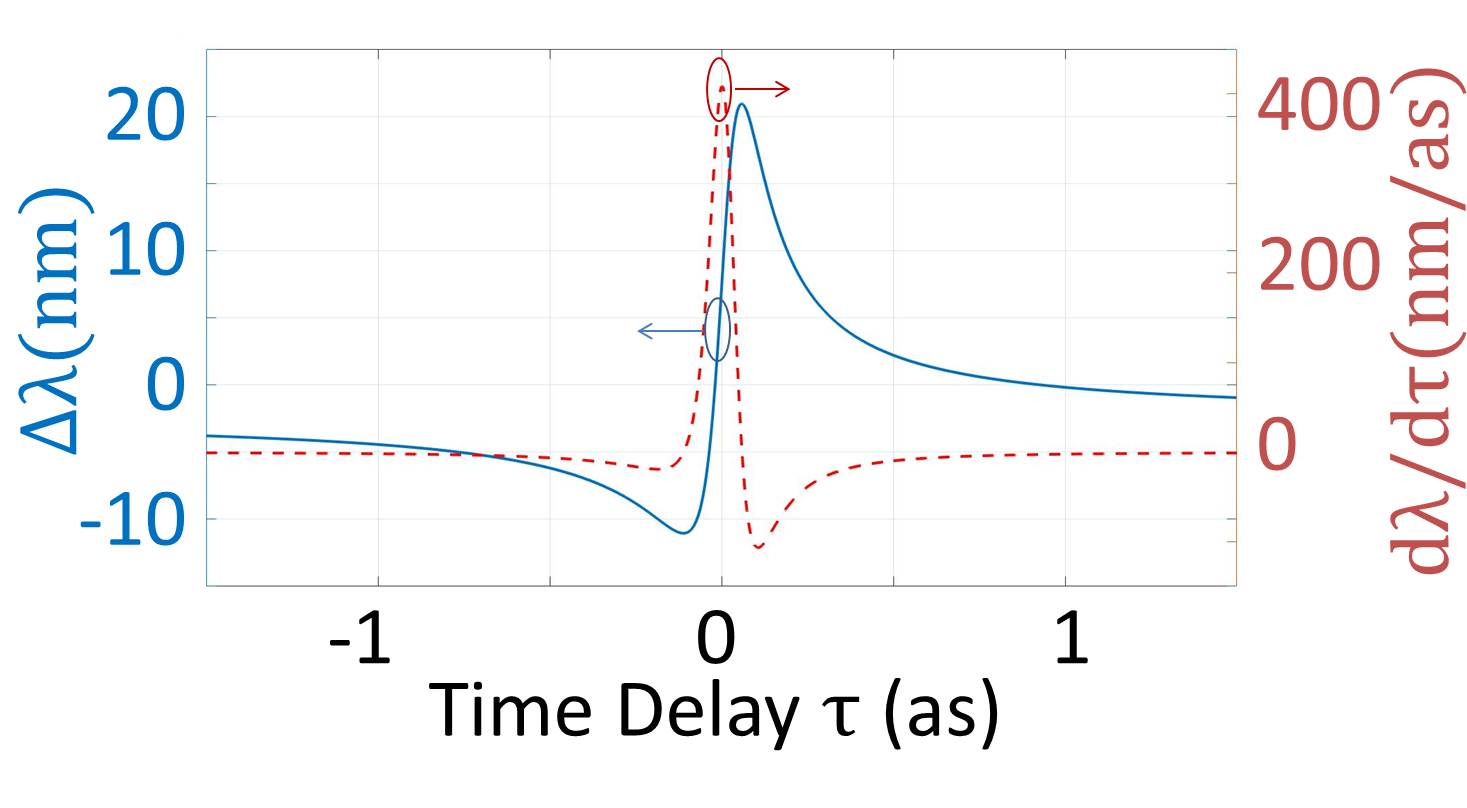}
    	\end{minipage}
    }
	\subfigure[]{
		\begin{minipage}[t]{0.47\linewidth}
			\centering
			\includegraphics[width=4.2cm]{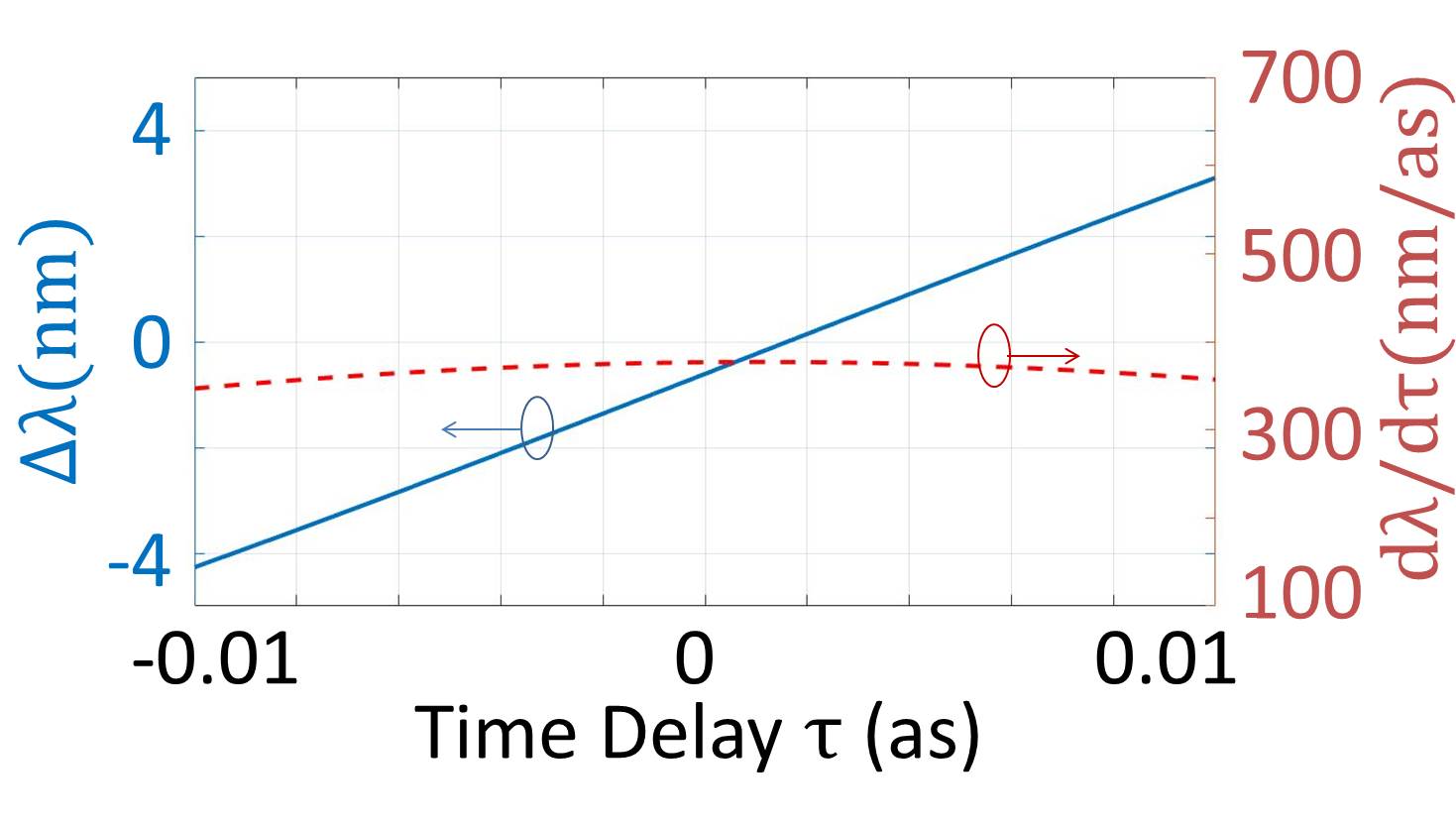}
		\end{minipage}
	}
    \subfigure[]{
    	\begin{minipage}[t]{0.47\linewidth}
    		\centering
    		\includegraphics[width=4.2cm]{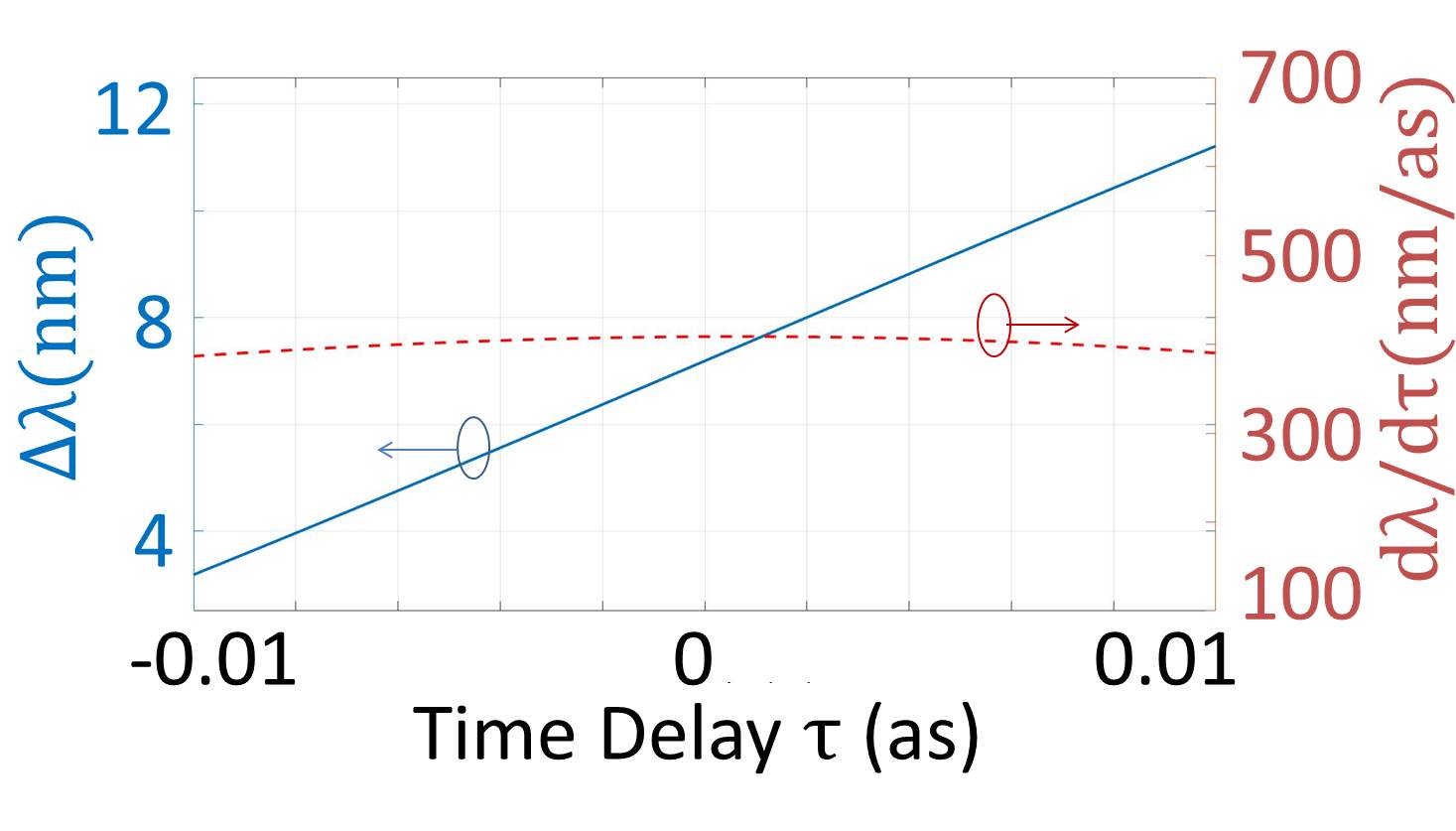}
    	\end{minipage}
    }
	\caption{(Color on line). Numerical simulation of DWVA scheme with $\epsilon = -0.01$ for Gaussian (a,c,e) and non-Gaussian (b,d,f) initial pointer state with comparison on their sensitivities (g): (a) Initial wavelength spectrum (Gaussian) with mean wavelength of 1540nm and line width of 25nm. (b) Initial wavelength spectrum (non-Gaussian) with mean wavelength value of 1540nm, which is captured from a light beam generated by SLD that passing through a polarizer. (c) Mean wavelength shift to Gaussian initialization and (e) zoom-in of the ultimate sensitive area. (d) Mean wavelength shift to non-Gaussian initialization and (f) zoom-in of the ultimate sensitive area. 
	%(g) The mean wavelength shift rates at $\tau=0$, for non-Gaussian pointer (red dash) and Gaussian pointer (blue solid).
    }
	\label{fig:simulation}
\end{figure}

For numerical analysis, we first stay with the Gaussian assumption, and then turn to discussions on non-Gaussian situation. For convenience, we use wavelength $\lambda$ instead of $\omega$ through the connection $\lambda = \frac{2\pi c}{\omega}$. The Gaussian wavelength spectrum is given by $P_G(\lambda) = \frac{1}{\sqrt{2\pi}\sigma_{\lambda}}\exp[-\frac{(\lambda-\lambda_0)^2}{2\sigma_{\lambda}^2}]$, where $\lambda_0$ and $\sigma_{\lambda}$ denote the mean wavelength and line width respectively, here we set $\lambda_0=1540nm$ and $\sigma_{\lambda}=25nm$. According to Eq.(\ref{eq:dwva}), at $\tau=0$ we can achieve the highest sensitivity, where the mean wavelength shift (MWS) corresponding to $\tau$ is given by
\begin{equation}\label{eq:shift_lam}
\delta\lambda \simeq \frac{4\pi c}{\epsilon}\tau.
\end{equation}
Interestingly, Eq.(\ref{eq:shift_lam}) is irrelevant to $\lambda_0$ and $\sigma_p
$. Numerical simulation results of MWS ($\delta\lambda$) and mean wavelength shift rate (MWSR) ($d\lambda/d\tau$) are shown in Fig.\ref{fig:simulation}(c) and (e) with assuming $\epsilon=-0.01$. In the most sensitive working area, MWSR up to $400nm/as$ can be achieved. Suppose the wavelength resolution is 0.01nm, the corresponding sensitivity of measuring $\tau$ is $2.5\times10^{-5}as$.

\section{Experiment}
\begin{figure}[!h]
	\centering
	\resizebox{8cm}{!}{
		\includegraphics[width=1.0\textwidth]{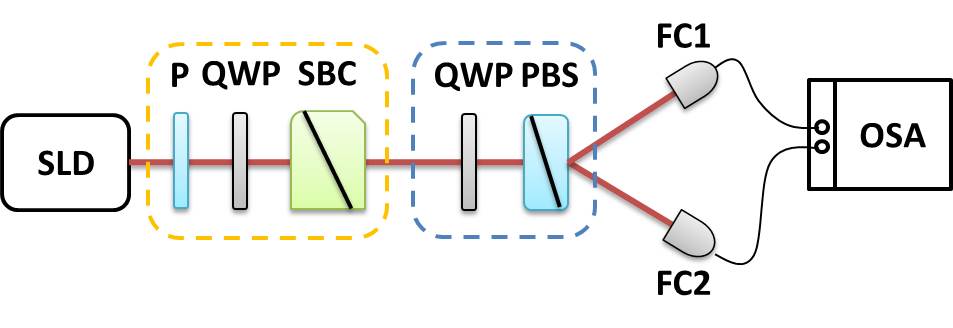}}
	\caption{(Color on line) Experimental setup for longitudinal phase change measurement using DWVA. SLD: super-luminescent diode, P: linear polarizer with optical axis at the angle of $\pi/4$, QWP: quarter wave plate with optical axis at the angle of $\pi/4 + 2\epsilon$ (former) and $0$ (later), SBC: Soleil-Babinet compensator, PBS: polarizing beam splitter splitting polarizations on directions of $\pi/4$ and $3\pi/4$, FC: fiber coupler, OSA: optical spectrum analyzer.
	}
	\label{fig:experiment}
\end{figure}

The experimental setup is shown in Fig.\ref{fig:experiment}. A light beam with broadband spectrum is generated by the super-luminescent diode (SLD) (Connet Fiber Optics Co.\cite{connet}, ordered for customization). A linear polarizer (Meadowlark, UHP-100-VIS-AR), a quarter wave plate (QWP, Thorlabs AQWP10M-1600) and a Soleil-Babinet compensator (SBC, Thorlabs SBC-IR) consist of the initial state preparation according to Eq.(\ref{eq:psi_in_final}). Here, the polarizer modulates a polarization state $(|H\rangle+|V\rangle)/\sqrt{2}$, and the optical axes of QWP and SBC are tilted by angles of $\epsilon/2$ and $\pi/4$ with respect to the polarizer. A longitudinal phase change of $\omega(\tau_0+\tau)$ between polarizations $H$ and $V$ is modulated by the SBC, where $\tau_0=\lambda_0\epsilon/2\pi c$ is fixed for initial state modulation, and $\tau$ is the parameter to be measured in the experiment. Afterward, the final state is postselected by adding another QWP together with a polarizing beam splitter (PBS, Thorlabs CCM1-PBS254), of which the axes at angles of $\pi/4$ and $0$ with respect to the polarizer, respectively. Finally, the lights getting out from two output ports of PBS are simultaneously sent to the optical spectrum analyzer through a $1\times4$ optical switcher (Yokogawa, AQ6370D and AQ2200-411).

Practical imperfections in experiment should be taken into consider. In particular, the Gaussian assumption of initial spectrum appearing in many WVA schemes analysis\cite{Howell2010,Brunner2010,Li2011} is hard to realized in practice. For instance, the shape of wavelength spectrum highly depends on the twisting optical fiber and other imperfect optical components used in experiment. Fortunately, DWVA could be available with non-Gaussian initial states. To show this, we perform the numerical simulation by applying a realistic wavelength spectrum as initial, which is captured by the light came out from SLD and passed through a polarizer (see Fig.\ref{fig:simulation}(b)). The MWS and MWSR are calculated and shown in Fig.\ref{fig:simulation}(d) and (f) respectively. 
Similar to the Gaussian case, the highest sensitivity is achieved at $\tau = 0$, with a non-zero bias of $\delta\lambda_{bias} \approx 7.186nm$. After calibration, we can use the following formula for estimating the parameter of interest:
\begin{equation}\label{eq:shift_lam_modified}
\tau = (\delta\lambda-\delta\lambda_{bias})/\frac{d\lambda}{d\tau}|_{\tau=0}.
\end{equation}
By setting $\epsilon=-0.01$, the highest MWSR that can be achieved at $\tau=0$ is $408 nm/as$. Suppose the wavelength resolution is 0.01nm, the sensitivity of measuring time-domain longitudinal phase change is $2.45\times10^{-5}as$, which is similar to the Gaussian case.
%\begin{widetext}
\begin{figure}[!h]
	\subfigure[]{
		\begin{minipage}[t]{0.8\linewidth}
			\centering
			\includegraphics[width=7cm]{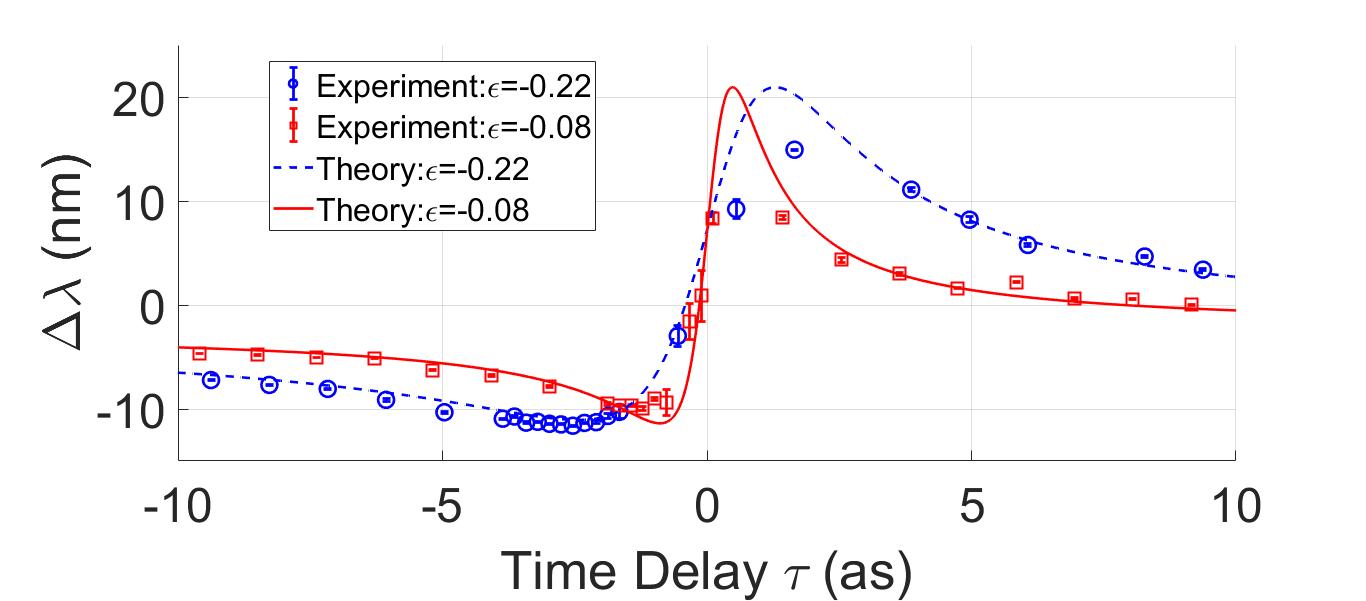}
		\end{minipage}
	}
	\subfigure[]{
		\begin{minipage}[t]{0.8\linewidth}
			\centering
			\includegraphics[width=7cm]{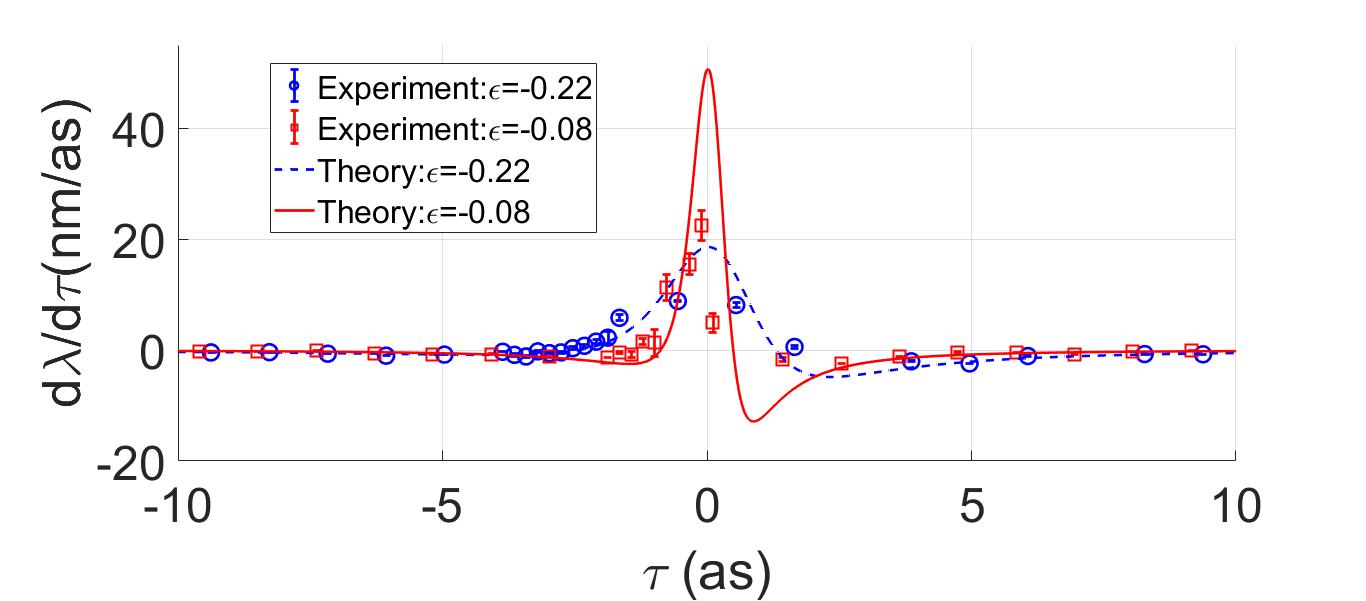}
		\end{minipage}
	}
	\subfigure[]{
		\begin{minipage}[t]{0.8\linewidth}
			\centering
			\includegraphics[width=7cm]{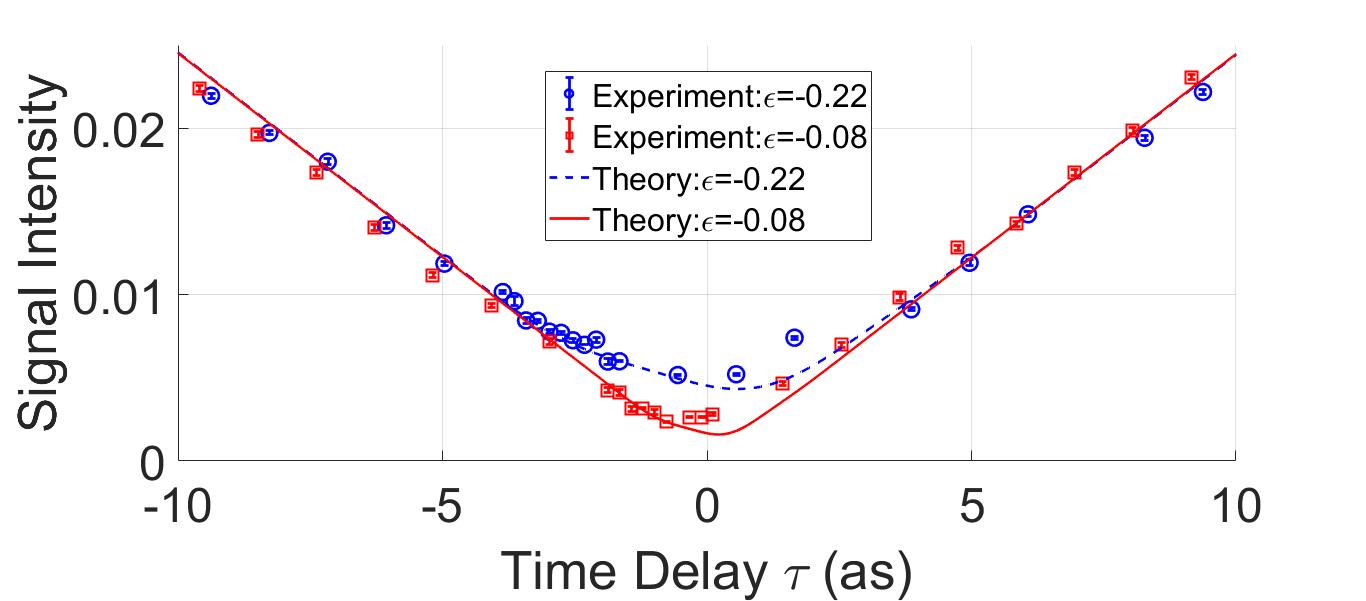}
		\end{minipage}
	}
	\caption{(Color on line). Experimental results in company with theory expectations of (a) Mean wavelength shift, (b) mean wavelength shift rate, and (c) signal intensity are shown, with setting $\epsilon=-0.08$ (red solid lines for simulations and red squares for experimental results) and $\epsilon=-0.22$ (blue dash lines for simulations and blue circles for experimental results). Modulation on $\epsilon$ is realized by tilting the optic axis of quarter-wave plate by a tiny angle, which is approximately $-2.3^o$ for $\epsilon=-0.08$ and $-6.2^o$ for $\epsilon=-0.22$.}
	\label{fig:exp_result}
\end{figure}
%\end{widetext}
%

The experimental results are shown in Fig.\ref{fig:exp_result}, with $\epsilon$ setting to be $-0.08$ and $-0.22$. Higher sensitivity could be achieved with smaller $\epsilon$, however in this case the range of the most sensitive working area would be narrower than the modulating resolution of SBC, behavior inside this area can not be revealed. We measured the MWS, MWSR and signal intensity, finding that experimental results fit well with the theoretical expectations. In our experiment, a MWSR of over $20(nm/as)$ is observed. Suppose the wavelength resolution is 0.01nm, sensitivity up to $5\times10^{-4}$as is achieved in our experiment. On the other hand, the minimum signal intensity in proportion to input is at the level of $10^{-3}$, which is 3 orders of magnitude higher than that of BWVA under identical conditions. 

\section{Conclusion} 
In summary, we propose the DWVA with two significant advantages over the other WVA approaches. First, DWVA can achieve sensitivity by several orders of magnitude higher than standard WVA without sacrificing signal intensity. Second, DWVA looses the requirement of using strictly Gaussian initial pointer state, which reduces the implementation complexity and makes it more compatible with practical imperfections. Taking these advantages into account, we perform a proof-in-principle experiment, where a tiny longitudinal phase change is converted to a remarkable mean wavelength shift by using DWVA. In our experiment, a mean wavelength shift rate up to over 20nm/as is observed, which indicates a sensitivity of $5\times10^{-4}$as can be achieved with 0.01nm wavelength resolution. 
Moreover, to evaluate the technical limit of DWVA under varied kinds of practical imperfections, such as effects of background noises and spectrometer resolution, further analysis based on signal-to-noise ratio or other figure of merit is required and currently under investigation\cite{Huang2019}. 
Finally, the idea of DWVA is not restricted to longitudinal phase change measurement but can also be applied to measure other kinds of small effects, providing a new method for precision metrology. 

\begin{acknowledgments} 
This work is supported by National Natural Science Foundation of China (Grant No. 61701302 and 61631014).
\end{acknowledgments}

\end{document}